\documentclass[12pt]{article}
\usepackage{graphicx}
\usepackage{amsbsy}
\def \bea{\begin{eqnarray}}
\def \beq{\begin{equation}}
\def \eea{\end{eqnarray}}
\def \eeq{\end{equation}}
\def \PDG{Particle Data Group, K. Hagiwara {\it et al.},
Phys.~Rev.~D {\bf 66}, 010001 (2002)}
\begin{document}

\Large

\centerline {\bf Synchrotron Radiation at Radio Frequencies}
\centerline {\bf from Cosmic Ray Air Showers~
\footnote{Enrico Fermi Institute preprint EFI-02-91, astro-ph/0211273.}}
\normalsize
\bigskip

\centerline {Denis A. Suprun$\,^a$~\footnote{d-suprun@uchicago.edu}, Peter
W. Gorham$\,^b$~\footnote{gorham@phys.hawaii.edu}  and
Jonathan L. Rosner$\,^a$~\footnote{rosner@hep.uchicago.edu}}
\vspace{0.5cm}
\centerline{\it $^a$ Enrico Fermi Institute and Department of Physics}
\centerline{\it University of Chicago, 5640 S. Ellis Avenue, Chicago, IL
60637}
\vspace{0.2cm}
\centerline{\it $^b$ Department of Physics and Astronomy, University of
Hawaii at Manoa}
\centerline{\it 2505 Correa Rd., Honolulu, HI 96822}
\bigskip
\bigskip

\centerline{\bf ABSTRACT}
\begin{quote}
We review some of the properties of extensive cosmic ray air showers
and describe a simple model of the radio-frequency radiation generated
by shower electrons and positrons as they bend in the Earth's magnetic
field. We perform simulations by calculating the trajectory and radiation
of a few thousand charged shower particles. The results are then
transformed to predict the strength and polarization of the
electromagnetic radiation emitted by the whole shower.

\end{quote}

\section{Introduction}

In the next few years the giant Pierre Auger air shower array will start
operating in the Southern Hemisphere, to be followed by a Northern Hemisphere
companion. The area covered by these arrays will be large enough to allow
detection of showers with energies higher than $10^{20}$~eV. The radiation
properties of such energetic showers differ from those with energies up to
$10^{17}$~eV that were routinely studied by smaller arrays. Particular
attention should be paid to the fact that very energetic air showers develop
to their maximum not far from the Earth's surface. While the typical altitude
of the maximum of a $10^{17}$~eV shower is located at
625~g/cm$^2$~\cite{Hires}, i.e.\ at 4~km elevation, a $10^{20}$~eV shower
maximum approaches $1$~km above sea level. This is about the same elevation as
that of many ground detectors.
The maxima of the most energetic showers will not be observed overhead but
rather from the side. The differences in the position and the distance to the
shower maxima affect the intensity and polarization of the radiation
emitted by these showers.

Many studies are concentrated on understanding the coherent Cherenkov 
radio emission from the excess charge in high energy showers developing
in dense media~\cite{AMZ,BR,GS1,GS2}. In air showers, however, another 
radiation
mechanism dominates the Cherenkov radiation. It is associated with the
acceleration that charged particles of the shower experience in the Earth's
magnetic field~\cite{KL,SC}.
This mechanism would occur even if there were no excess charge in air 
showers.
Electrons and positrons bend in the opposite directions in a magnetic field.
However, the opposite signs of their accelerations are cancelled by the 
opposite signs
of the electric charges. As a result, the electromagnetic radiation from both
particles of an electron-positron pair is coherent, becoming the main source 
of the air shower radio emission~\cite{Allan}. 
In this paper we develop a simple radiation model which can
be applied to most geometries relevant to extensive air showers.

In Section~II we consider the electric field produced 
by a single accelerated charged particle. Previous experimental 
studies of the shower radiation strength are discussed in Section~III. 
Section~IV reviews shower pancake 
structure and restrictions it imposes on the choice of the
radiation model. Coherence of shower particle 
radiation is considered in Section~V, while Sections~VI and~VII give
details and results of the Monte Carlo simulations. We summarize in 
Section~VIII. Details of the radiation calculations
and symmetries of the radiation patterns are given in an Appendix.

\section{Radiation from accelerated charges}

Numerous experiments have confirmed acceleration of charged particles in the
Earth's magnetic field as the main source of the electromagnetic radiation
from an extensive air shower (see in particular~\cite{RS} and several other
articles in \cite{RADHEP}).  For a $10^{17}$~eV shower with a maximum high
above the Earth the radiation of each gyrating particle can be understood as
familiar synchrotron radiation~\cite{FG}. Even in this case the significant
lateral and
longitudinal spread of the shower particles hinders simple analytical
calculation of the total electromagnetic radiation.
For a $10^{20}$~eV shower the distance between the maximum and an antenna
detecting the radio-frequency radiation could be comparable to the radiation
length of an electron (36.7~g/cm$^2$, or 330~m at an altitude of $1$~km).
This means that the displacement of a radiating particle is significant and
the direction from it towards the antenna appreciably changes during the
motion. As a result, the usual synchrotron formulae are not easily
applicable even for a single electron.
Instead, we start from the underlying
formula for a radiating particle~\cite{Jackson,ZHS}: 
\beq {\bf E}({\bf x},t_a)=
\frac{e \mu}{4 \pi \epsilon_0 } \,\left[\frac{{\bf
n}-n\boldsymbol{\beta}} {\gamma^2 
|1-n\boldsymbol{\beta}\cdot{\bf n}|^3\,l^2}\right]_{\rm ret} + 
\frac{e \mu}{4 \pi \epsilon_0
c}\,\left[\frac{{\bf n}\times\left[({\bf
n}-n\boldsymbol{\beta})\times\dot{\boldsymbol{\beta}}\right]}
{|1-n\boldsymbol{\beta}\cdot{\bf n}|^3\,l}\right]_{\rm ret}
\label{elfield} \eeq which is correct regardless of the distance
to the antenna. In this formula $\boldsymbol{\beta}$ is the
velocity vector in units of $c$,
$\dot{\boldsymbol{\beta}}=d\boldsymbol{\beta}/dt$ is the
acceleration vector, divided by $c$, ${\bf n}$ is a unit vector
from the radiating particle to the antenna, and $l$ is the
distance to the particle. $\mu\approx1$ denotes the relative
magnetic permeability of air, $n$ the index of refraction. The
square brackets with subscript ``ret" indicate that the quantities in
the brackets are evaluated at the retarded time, 
not at the time $t_a$ when the signal arrives at the antenna.

The first term decreases with distance as $1/l^2$ and represents a
boosted Coulomb field. It does not produce any radiation. The
magnitudes of the two terms in Eq.~(\ref{elfield}) are related as
$1/(\gamma^2 l)$ and $|\dot{\boldsymbol{\beta}}|/c$.
The characteristic acceleration of a 30~MeV electron ($\gamma\approx60$) of
an air shower in the Earth's magnetic field
($B\approx0.5$~Gauss) is $|{\bf a}|=ecB/(\gamma m) \approx
4.4\cdot10^{13}$~m/s$^2$.
Even when an electron is as close to the antenna as 100~m,
the first term is two orders of magnitude
smaller than the second and can be neglected. The second term falls as $1/l$
and is associated with a radiation field. It describes the electric field of
a single radiating particle for most geometries relevant to extensive
air showers. It can be shown~\cite{FW} to be proportional to the apparent
angular acceleration of the charge up to some non-radiative terms that are
proportional to $1/l^2$. This relation is referred to in the literature as
``Feynman's formula.'' Sometimes it is used for the purposes of Monte Carlo
simulation~\cite{Hough}. For us, however, the explicit expression for the 
electric field in terms of velocity and acceleration in 
Eq.~(\ref{elfield}) turns out to be more convenient.

\section{Experimental observations}

The collaboration of H. R. Allan at Haverah Park in England
\cite{Allan} studied the dependence of radiation strength on primary energy
$E_p$, perpendicular distance $R$ of closest approach of the
shower core, zenith angle $\theta$, and angle $\alpha$ between
the shower axis and the magnetic field vector.
Their results indicate that the electric field strength per unit of
frequency, ${\cal E}_\nu$, could be expressed as
\beq
\label{eqn:E}
{\cal E}_\nu = s_r\,\frac{E_p}{10^{17}\,{\rm~eV}}\,\sin\alpha\,\cos\theta\,
\exp \left( - \frac{R}{R_0(\nu, \theta)} \right)~~~\mu{\rm
V}/{\rm m}/{\rm MHz}~~~,
\eeq
where $R_0$ is an increasing function of $\theta$, equal (for example) to
$(110 \pm 10)$ m for $\nu = 55$~MHz and $\theta < 35^\circ$. The equation is
valid for $E_p$ between $10^{17}$ and $10^{18}$~eV and for $R$ less than
300~m. Originally, the value of the calibration factor was determined to be
$s_r=20$. It was subsequently updated in~\cite{Atrash} to
yield field strengths approximately 12 times weaker, corresponding to
$s_r=1.6$, while observations in the
U.S.S.R. gave field strengths approximately 2.2 times weaker ($s_r=9.2$).
The latter two values of $s_r$ were obtained with approximately 10 times
higher
statistics and are a better indication of the shower radiation strength.
Still, some question persists about the magnitude
of the effect, serving as an impetus to further measurements.

We will describe a simple model of air shower radiation and then test if it
gives radiation strength corresponding to values of $s_r$ in the range from
1 to 10. We will take into account some characteristics of the Haverah Park
site, namely, its elevation (220~m), the Earth's magnetic field strength
($B=0.49$~Gauss) and inclination ($\gamma_d=68^\circ$).

\section{Radiation model and pancake structure}

We begin by making 
a simplifying assumption that radiation only comes
from the shower maximum and lasts as long as particles travel through one
radiation length ($\sim$1~$\mu$s). The total number $N_e$ of charged
particles
at the shower maximum is approximately 2/3 per GeV
of primary energy~\cite{PDG}. We will also assume equal numbers of 30~MeV
positrons and electrons among charged particles, neglecting an admixture of
muons, an excess of electrons and variations of energies between the
particles. Such a simple model serves as a precursor to a future full scale
shower development and radiation simulation similar to a study of the
properties of electromagnetic showers in dense media performed in~\cite{ZHS}.
Now we review lateral and longitudinal particle distributions in the shower
pancake.

Lateral particle density $\rho_e$ is parameterized by the age parameter $s$ 
of the shower ($s=1$ for the shower maximum) and the Moli\`ere radius
$r_m$~\cite{lateral,G,NK}:
\beq
\label{eqn:lateral}
\rho_e=K_N\,\left(\frac{r}{s_m r_m}\right)^{s-2}\,
\left(1+\frac{r}{s_m r_m}\right)^{s-4.5},
\eeq
where
\beq
K_N=\frac{N}{2\pi s_m^2 r_m^2}\,
\frac{\Gamma(4.5-s)}{\Gamma(s)\Gamma(4.5-2s)}  ,
\eeq
$\Gamma$ is the gamma function, $r$ the distance from the shower axis, $N$
the total number of charged particles, and $s_m=0.78-0.21s$. 
The Moli\`ere
radius for air is approximately given by $r_m=74\,(\rho_0/\rho)$~m, with 
$\rho_0$ and $\rho$ being the air densities at sea level and the altitude 
under consideration, respectively.

As a shower travels towards the Earth and enters denser layers of the
atmosphere, the age parameter increases while the Moli\`ere radius drops. 
Both processes affect the spread of the
lateral distribution. The influence of the age parameter appears to be
more significant. As it grows, the average distance of the shower
particles from its axis increases. This effect overcomes the influence of a
smaller Moli\`ere radius which tends to make the lateral distribution more
concentrated towards the axis. For a fixed age parameter $s$, however,
the Moli\`ere radius is the only quantity
that determines the spread of the lateral distribution.
At shower maximum ($s=1$) the average distance from the axis can be
calculated to be $(2/3) s_m r_m = 0.38\, r_m$.

The thickness of the shower pancake was also taken into account. It is
directly related to the average dispersion in arrival times of air shower
particles. At a distance $r$ from the axis it is described
in~\cite{thickness} by:
\beq
\sigma_t=\sigma_{t_0}\,\left(1+\frac{r}{r_t}\right)^b,
\label{eqn:longitudinal}
\eeq
where $\sigma_{t_0}=1.6$~ns, $r_t=30$~m and $b=(2.08\pm0.08)-
(0.40\pm0.06)\sec\theta+ (0\pm0.06)\log(E_p/10^{17}~{\rm eV})$, for $r<2$~km,
$10^{17}<E_p<10^{20}$~eV and $\theta<60^\circ$. This empirical formula
reflects an increase of the pancake thickness at greater distances from the
shower axis. It was derived from measurements performed at an altitude of
1800~m in New Mexico. For vertical showers it gives $b=1.68$.

For those showers that reach their maximum at 1800~m above sea level, one can
use the lateral distribution~(\ref{eqn:lateral}) to calculate the
average $\sigma$ of the shower pancake thickness at this altitude:
$\sigma_0=8.4$~ns.  For simplicity, we will assume that the longitudinal
distribution is independent of distance from the axis and has a constant
dispersion $\sigma_0$.

The shape of the longitudinal distribution is defined by multiple scattering
of electrons in the atmosphere. One can expect a sharp increase of particle
number at the bottom of the pancake where virtually unscattered high energy
particles are concentrated, followed by a slow decay at the top of the
pancake as slower electrons increasingly lag behind. This presumption was
confirmed in~\cite{shape} with the shape function being measured to be
proportional to $t^2 \exp(-t/\tau)$.

Unfortunately, the thickness of the pancake was not measured at various
altitudes above sea level. We would like to know the longitudinal
distribution at shower maximum, but the pancake thickness is known only at
1800~m. To circumvent this problem, we choose a shower of such energy that it
develops to its
maximum at about 1800~m altitude. Locations of the maxima of $10^{17}$ to
$10^{18.3}$~eV showers were measured by the HiRes/MIA hybrid detector to lie
higher in the atmosphere~\cite{Hires}.
They also obtained evidence of change from a heavy to a light composition
in this energy region. If the elongation rate remains constant for even
higher energies that would indicate that the primaries become exclusively
protons at about $10^{19}$~eV. Simulations using the SIBYLL high energy
hadronic interaction model show~\cite{Hires,Pryke} that $10^{19}$~eV protons
produce showers with maxima at 820~g/cm$^2$. This corresponds to 1800~m
altitude above sea level for vertical showers. For these showers
Eq.~(\ref{eqn:longitudinal}) gives a longitudinal distribution of charged
particles exactly at shower maximum. Only for these showers do we know both
lateral and longitudinal distributions at their maxima. 

\begin{figure}[p]
\centerline{\includegraphics[width=.85\textwidth]{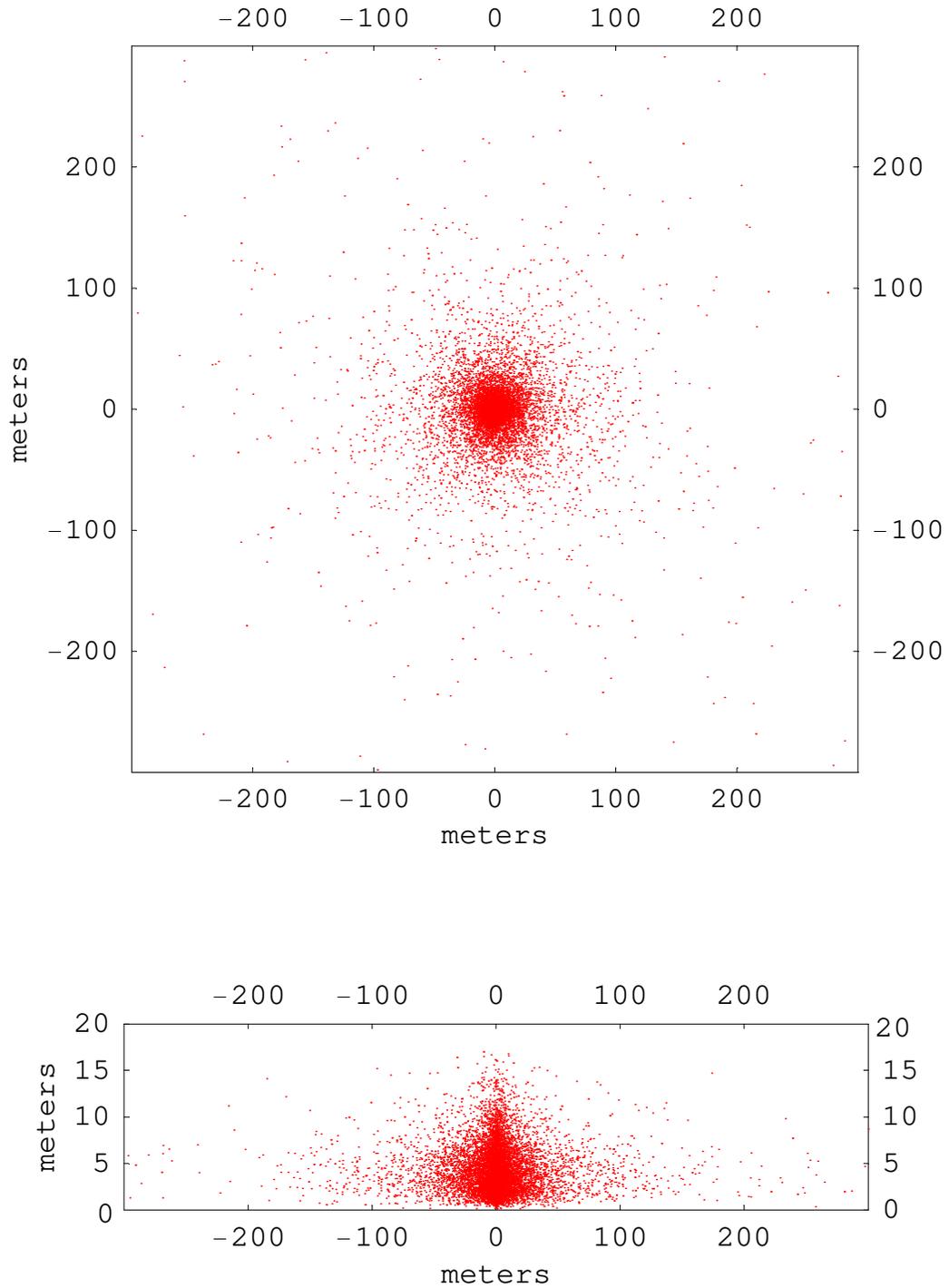}}
\caption{Top and side view of a shower pancake of a vertical $10^{19}$~eV 
shower at its maximum. The distributions of 10000 pancake
particles are shown. The average radial spread is 33~m.
The average delay of pancake particles is 4.4~m behind the central pair 
at the bottom of the pancake.}
\label{fig:profile}
\end{figure}

Thus, in the model we will use a vertical $10^{19}$~eV shower which develops
to its maximum of approximately $0.33 \cdot 10^{10}$ electrons and $0.33 \cdot
10^{10}$ positrons at 1800~m above sea level. The lateral distribution of the
pancake is described by Eq.~(\ref{eqn:lateral}) while the longitudinal
distribution follows a $t^2 \exp(-t/\tau)$ shape. The time constant $\tau$ 
was chosen to be equal to $\sigma_0/\sqrt3$. This ensures that the
dispersion of the longitudinal distribution is $\sigma_0=8.4$~ns,
or $2.5$~m. The pancake profile described in this Section is shown in 
Fig.~\ref{fig:profile}.

\begin{figure}[p]
\centerline{\includegraphics[width=.9\textwidth]{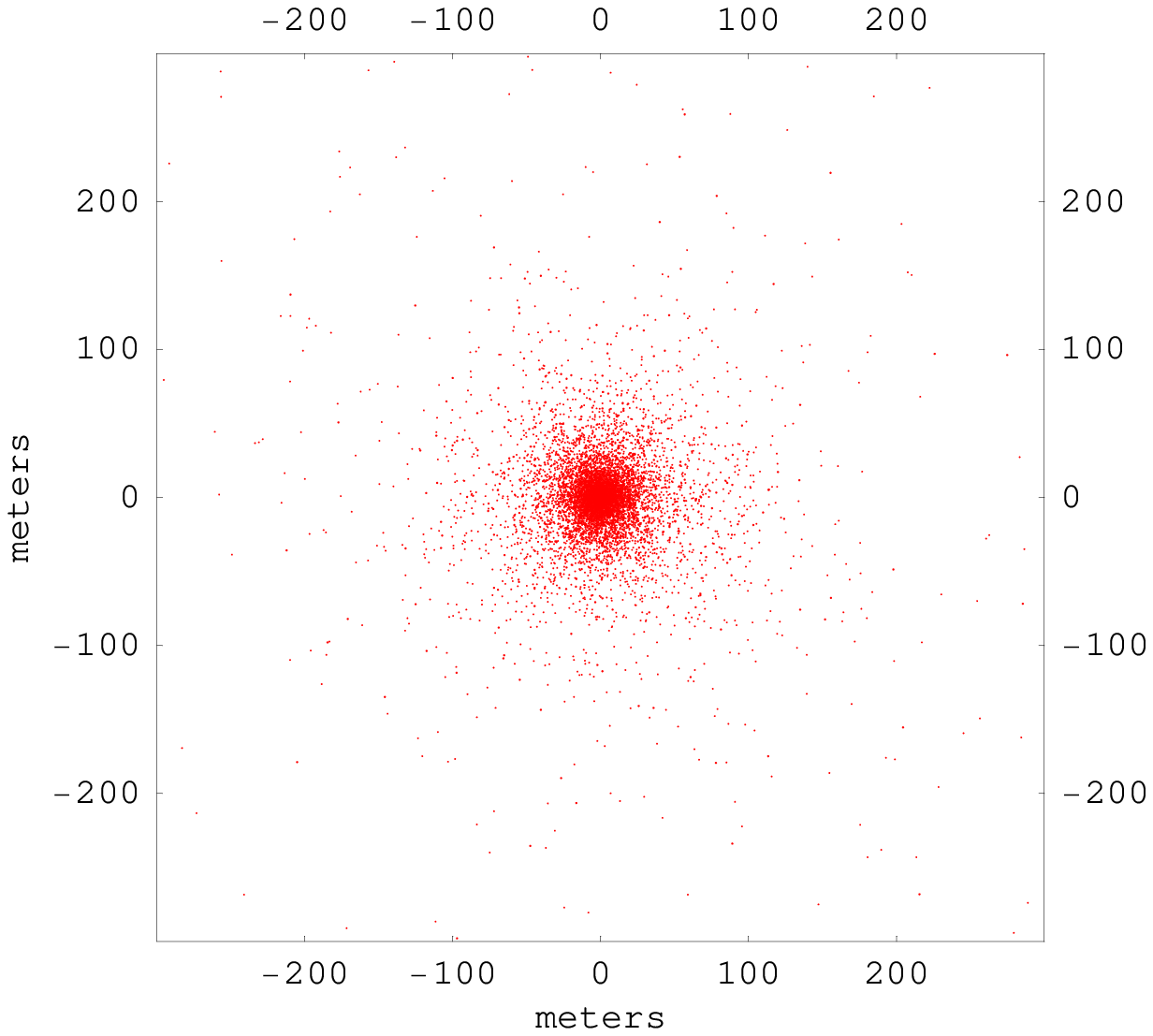}}
\end{figure}

\section{Coherence of pancake radiation}

The question remains if Eq.~(\ref{eqn:E}) correctly describes the radiation
from a $10^{19}$~eV shower. One does expect a linear growth of the pulse
amplitude as the number of radiating particles increases with shower energy.
The coherence of the emitting particles is an underlying assumption. It is
valid for radiation at RF wavelengths of several meters coming from the
main bulk of the particle swarm concentrated around the shower axis.
This simple argument is modified by two competing effects.

A greater penetration of the atmosphere by a more energetic shower means
that the shower develops to its maximum at a lower altitude with a higher
air density.
The Moli\`ere radius $r_m$ is smaller at this altitude and
Eq.~(\ref{eqn:lateral}) predicts a more concentrated lateral distribution of
the shower particles. As a result, the RF radiation of the shower should
become more coherent and stronger.
Thus, the pulse amplitude should increase more rapidly than linearly with
primary energy.

\begin{figure}[t]
\centerline{\includegraphics[width=.8\textwidth]{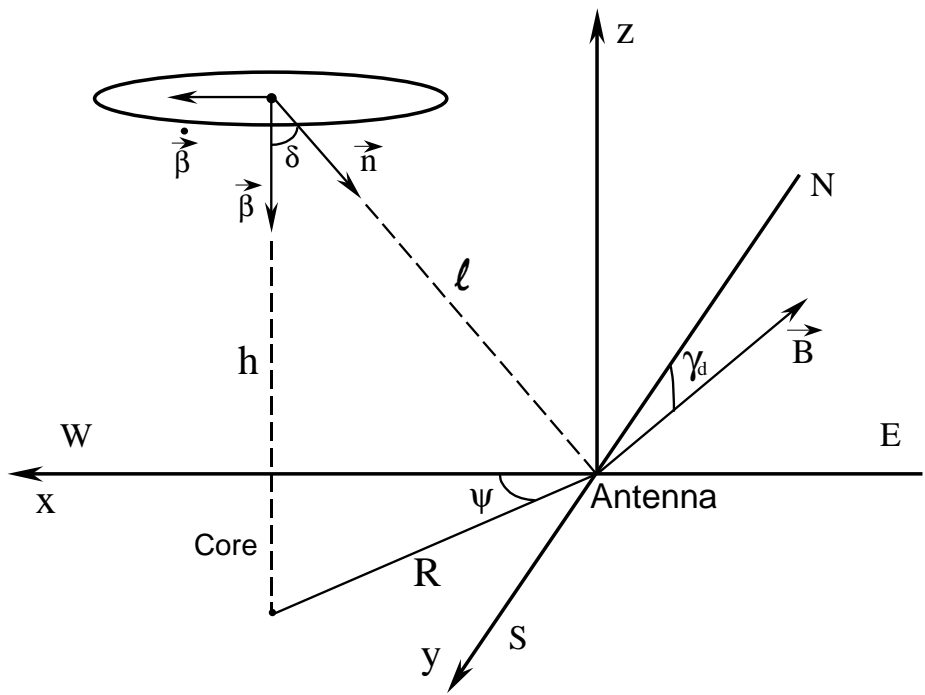}}
\caption{Geometry of a vertical shower. Axes relative to  antenna
are $x$ (magnetic West), $y$ (magnetic South) and $z$ (up).
Vector {\bf B} lies in the $yOz$ plane. The ellipse shows the plane
of a ``slice": an infinitely thin horizontal layer of the shower
pancake.}
\label{fig:geometry}
\end{figure}

However, this effect is offset by the fact that charged
accelerating particles radiate mostly in the direction of their
velocities.  For a fixed distance $R$ from antenna to 
the axis of a vertical shower, 
the angle $\delta$ between the vectors ${\bf n}$ and
$\boldsymbol{\beta}$ becomes larger (see Fig.~\ref{fig:geometry})
and the electric field smaller as the altitude decreases. This can
be seen directly from Eq.\ (\ref{elfield}). Considering small
angles $\delta$ and taking the index of refraction of air $n=1$
for simplicity, one can write its denominator as $(1-\cos\delta)^3
l \propto \delta^6 (R/\delta) \propto \delta^5$. The magnitude of
the numerator can be shown to be proportional to
$R^2/h^2$~\cite{NIM}, i.e., to $\delta^2$. As a result, the
electric field $E \propto \delta^{-3}$ and decreases at lower
altitudes. This should lead to a smaller pulse amplitude than
expected from a linear growth with shower energy.

The combination of different factors fortuitously leads to an overall linear
dependence of pulse amplitude on primary energy in the range between 
$10^{17}$
and $10^{18}$~eV, as shown in Eq.~(\ref{eqn:E}). As the primary energy
approaches $10^{19}$~eV, there is some evidence that the radiation strength
increases less rapidly than $E_p$~\cite{Allan}. Thus, we expect that for a
$10^{19}$~eV shower the calibrating factor $s_r$ in Eq.~(\ref{eqn:E}) shifts
down to the lower end of the 1 to 10 range.

\section{Monte Carlo details}

Calculation of the signal from $10^{10}$ particles is impossible with a
limited computing time. Instead, we start with a calculation of radiation
from an infinitely thin horizontal pancake slice located at 1800~m elevation.
$10^4$ electron-positron pairs were chosen randomly from this slice according
to the expected lateral distribution (see Fig.~\ref{fig:profile}, top panel).
The geomagnetic distortion of the axial symmetry of the pancake was not taken
into account.
The initial velocity of particles was
assumed to be vertical. Then, the trajectory, velocity and acceleration in
the Earth's magnetic field of each gyrating particle were calculated
according to Eqs.~(\ref{eqn:re}-\ref{eqn:ap}) (see Appendix). After taking
into account the delay associated with the propagation of the
electromagnetic
wave, we compute the electric field perceived by the ground based antenna
located at 220~m above sea level. The East-West electric field components
from different pairs are summed up to simulate the electromagnetic pulse
detected by the EW-oriented antenna.

\begin{figure}[t]
\centerline{\includegraphics[width=.75\textwidth]{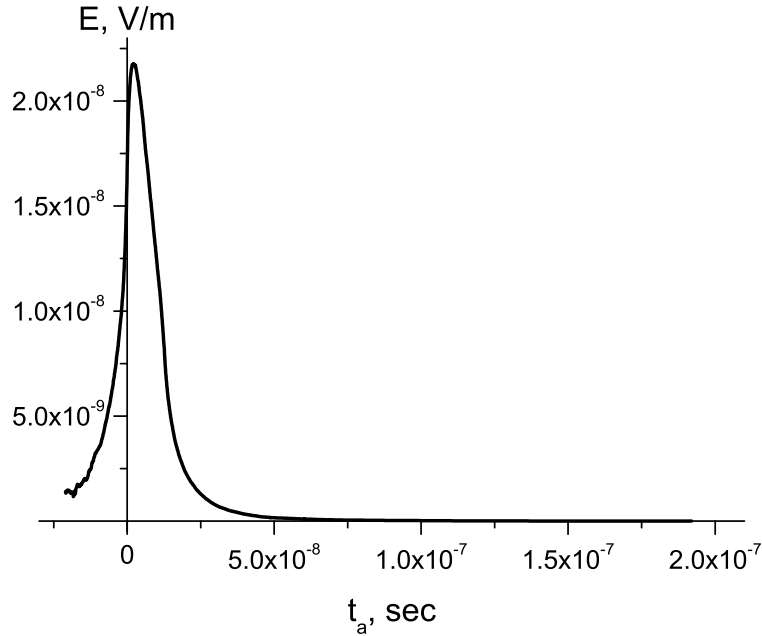}}
\caption{The EW component $E_{EW}$ of the electromagnetic pulse of $10^4$
radiating electron-positron
pairs distributed over a horizontal slice at 1800~m above sea level. The
axis of the slice is located 200~m South of the antenna. The time axis was
chosen in such a way that the pulse of a pair located at the axis of the
slice starts at time 0. The earliest possible arrival time is $-42$~ns and
comes from pairs located right above the antenna. 
Those pairs are too far from the axis 
(200~m being much larger than the average spread of $0.38\,r_m=33$~m) 
and the statistics for them is not sufficient.}
\label{fig:slice}
\end{figure}

In the process of motion charged particles become closer to the antenna.
As a result, the observed
pulse duration is much shorter than the real radiation time of the order of
1~$\mu$s. The radiation from different pairs does not arrive at the same time
but the short individual pulses overlap to form a ``slice" pulse. The
statistics is high enough for the particles around the shower axis and the 
pulse
amplitude becomes asymptotically proportional to the number of radiating
particles. That allows us to determine the electromagnetic pulse from an
arbitrary high number of particles in the slice by simple scaling. The pulses
from pairs that are further from the center overlap only slightly. As a
result, the net pulse from the distant particles is not smooth and displays
statistical jitter. It is not proportional to the number of particles
involved. With a limited computing time one cannot avoid this problem. We do
not amplify random fluctuations by scaling them to a higher statistics.
Instead, we try to determine the region of ``sufficient statistics". To do
that we take the ratio of the pulse produced by 10000 pairs and the pulse
produced by 5000 pairs. The time interval where this ratio is within 25\% from
2 is assumed to be the region of sufficient statistics and eligible for
scaling to a higher number of pairs. Fig.~\ref{fig:slice} shows the shape of
the slice pulse $E(t_a)$ in the region of sufficient statistics. The fraction
of particles that produce radiation in this time interval is 96.7\%. It is a
representative fraction of the total radiation. We will neglect the radiation
outside of this time interval.

\begin{figure}[p]
\centerline{\includegraphics[width=.6\textwidth]{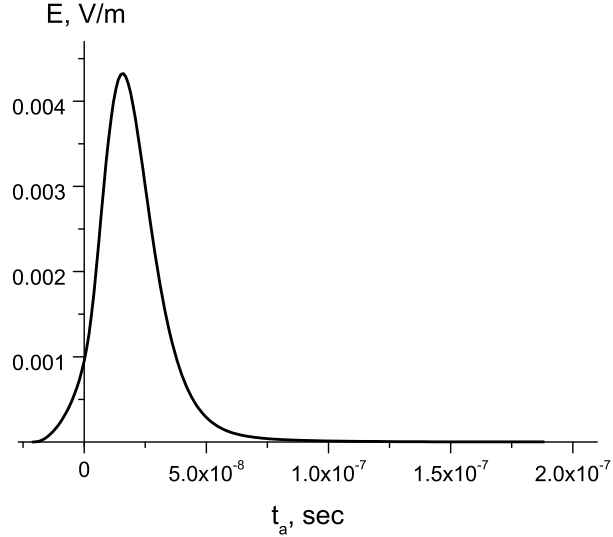}}
\caption{The EW component $E_{EW}$ of electromagnetic pulse of
$0.33\cdot10^{10}$ radiating
electron-positron pairs distributed over the thickness of the shower pancake
at 1800~m above sea level. The axis of the pancake is located 200~m South
of the antenna. The time axis was chosen in such a way that the pulse 
produced by a pair located in the axis at the bottom of the pancake starts 
at time 0.}
\label{fig:pancake}
\end{figure}

\begin{figure}[p]
\centerline{\includegraphics[width=.6\textwidth]{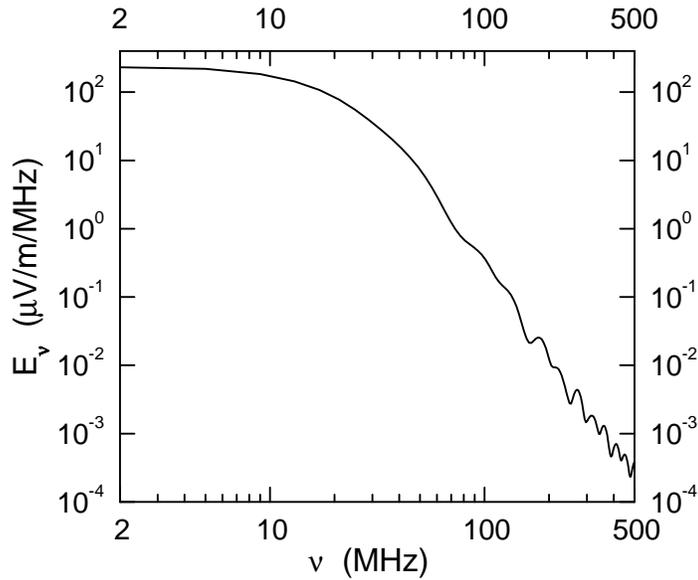}}
\caption{The Fourier transform of the electromagnetic pulse shown in
Fig.~\ref{fig:pancake}. The spectrum is very flat below 2~MHz.
The limited statistics of the model results in some jitter at 200$-$500~MHz.
The spectrum above 500~MHz is not shown because the statistics is 
not sufficient to make reliable predictions of the Fourier components at 
these high frequencies.}
\label{fig:FT}
\end{figure}

Now that the slice pulse is determined, we take into account that slices are
longitudinally distributed according to the $t^2 \exp(-t/\tau)$ formula,
$\tau=\sigma_0/\sqrt3$. The dispersion of their distribution
$\sigma_0=8.4$~ns corresponds to the distance of 2.5~m. It is very small
compared to the distance between the shower pancake and the antenna.
Therefore, each horizontal slice produces the same signal as the one shown in
Fig.~\ref{fig:slice}, only shifted in time. As a result, the problem of
calculating
the pancake pulse reduces to the integration of appropriately weighted and
shifted slice pulses. The result is scaled up for $0.33\cdot10^{10}$ pairs
and shown in Fig.~\ref{fig:pancake}. Fig.~\ref{fig:FT} gives the Fourier
transform $E_\nu$ of the pancake pulse.  The nonzero thickness of the pancake
translates into a longer pancake pulse in comparison with the slice pulse,
thereby limiting the main part of the radiation spectrum to the frequencies 
below 100~MHz.

\section{Results and discussion}

\begin{figure}[t]
\centerline{\includegraphics[width=.75\textwidth]{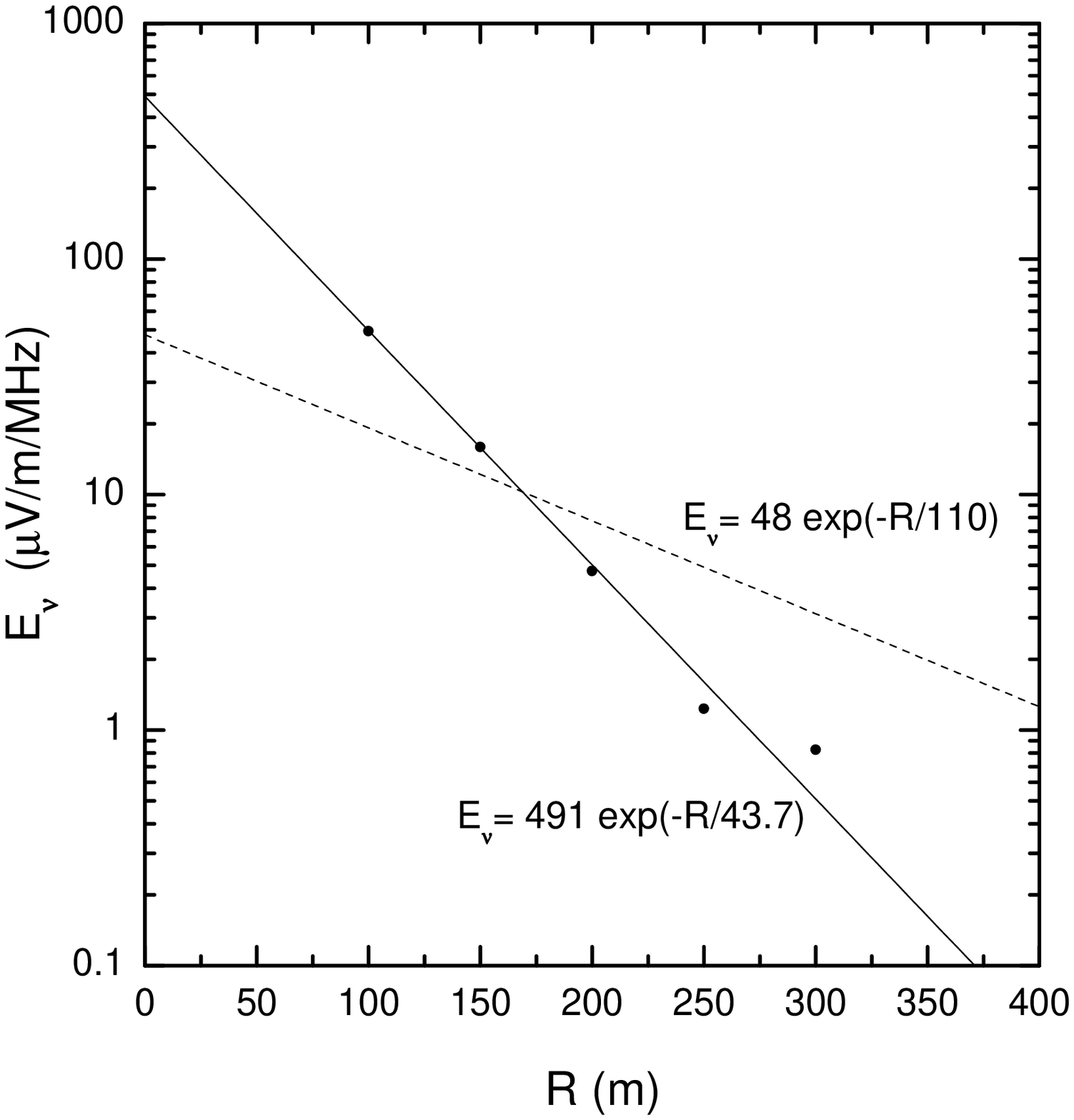}}
\caption{The East-West component of the field strength $|{\cal E}_{\nu EW}|$
at 55~MHz as a function of distance from the shower axis. The dashed line is 
the best fit of formula~(\ref{eqn:E}) to the Monte Carlo results. The 
calibrating parameter for the fit is $s_r=1.27$. The solid line is the best 
exponential fit with characteristic distance $R_0=43.7$~m. The statistical 
uncertainties from repeated Monte Carlo simulations are smaller
than sizes of the dots. Nevertheless, they were taken into account during 
fitting procedures.}
\label{fig:R_dep}
\end{figure}

\subsection*{Radio pulse spectrum.}

The spectrum (Fig.~\ref{fig:FT}) shows a clear power law falloff  
between 40--500~MHz, approximately $\sim1/\nu^{4.5}$.
This observation can be compared to the measurements~\cite{AC}
of the radiation spectrum that observed that the field strength 
$|{\cal E}_{\nu}|$ goes as $1/\nu$ between 1--500~MHz. However, these 
measurements were admittedly
inconclusive~\cite{Allan} because they effectively averaged the spectral 
distribution over different distances between antenna and shower core. At 
large distances $R$ from the axis the radio pulse becomes longer. In our 
model the main part of the signal is concentrated in the first 50~ns for 
$R=100$~m but increases to about 150~ns for $R=300$~m. This leads to a loss 
of high frequency components and the spectrum is expected to be steeper at 
larger core distances. The expectation that the spectrum shape may be 
different at different distances led to a conclusion that the measured 
$1/\nu$ spectrum is not of any fundamental significance but simply reflects 
the characteristics of the detecting array. One can only expect that for a 
fixed $R$ the spectrum  decreases with frequency. There is a general 
agreement between the predicted spectrum for $R=200$~m (Fig.~\ref{fig:FT})  
and this expectation.

It is worth noting that particles of different energies are present in a 
real shower pancake. Those with lower energies increasingly lag behind.
The lower the energy of a particle, the later its radiation arrives to the 
antenna. This radiation is also stronger for low energy particles experience 
higher accelerations. As a result, the pulse of the real pancake differs
from the one in Fig.~\ref{fig:pancake} in that its peak occurs at a
later time and, more importantly, its width is larger. The preponderance 
of low frequency components in the spectrum of the real pancake should be 
even more prominent than that of Fig.~\ref{fig:FT}.

\subsection*{Variation of field strength with distance to the antenna.}

We calculated electromagnetic pulses for the pancakes whose axes are located
at various distances $R$ South of the antenna. To compare with the Haverah
Park measurements~\cite{Allan} we show the variation of the East-West
component of the field strength $|{\cal E}_{\nu EW}|$ at 55~MHz with distance
$R$ (Fig.~\ref{fig:R_dep}). 

A power law fitting these results would rise too fast at distances
smaller than 100~m. It makes more sense to use an exponential fit.
We can fit formula~(\ref{eqn:E}) to the results of the Monte Carlo 
simulation. For $10^{19}$~eV vertical 
showers at the Haverah Park site, this formula becomes ${\cal E}_\nu = 
100s_r\cos\gamma_d\exp(-R/R_0)=37.5 s_r \exp(-R/110)$. The best fit 
is provided by the calibrating factor $s_r=1.27$. 
This value is indeed in the lower end of 1 to 10 range, as
expected for the $10^{19}$~eV shower. One can tell that even this simple
model is successful in predicting the order of magnitude of the
electromagnetic pulse emitted by the extensive air showers.

The best exponential fit, however, is provided not by Eq.~(\ref{eqn:E}) with
$R_0=110$~m, but by an exponential with characteristic decay distance
$R_0=43.7$~m. This may be a consequence of either the limited nature and 
simplicity of this model or the great experimental difficulties besetting 
the detection and calibrating radio pulses from the extensive air showers. 
A future full scale shower development simulation and a possible equipment 
of the Auger experiment with RF detecting antennas will clarify this issue.

\subsection*{Variation of field strength with angle $\psi$.}

Consider the frame centered at the antenna, with axis $Ox$ going
to  the magnetic West, $Oy$ to the South and $Oz$ directly up. The
initial velocity of all charged particles is assumed to be
vertical: $\boldsymbol{\beta}=(0,0,-1)$, while the initial
acceleration $\dot{\boldsymbol{\beta}}$ is parallel to $Ox$, or,
in other words, to the $(1,0,0)$ vector (see
Fig.~\ref{fig:geometry}). Electrons bend towards the magnetic West
and positrons towards the East. The electric fields from both
particles of an electron-positron pair are coherent; the opposite
signs of their accelerations are cancelled by the opposite signs
of the electric charges.

Let $\psi$ be the angle between
$Ox$ and the direction to the shower core, $R$ the distance to the
core, and $h$ the altitude of the radiating
particle above the antenna.
The denominator of the second term of Eq.~(\ref{elfield}) is independent
of $\psi$.
The numerator determines that, to leading
(second) order in $R/h$, the initial
electric field vector ${\bf E}$ received at the antenna lies in the
horizontal plane and is
parallel to $(\cos2\psi,\sin2\psi,0)$~\cite{NIM}:
\beq
{\bf E} \ \| \ (\cos2\psi,\sin2\psi,0)
\label{eqn:Eparallel}
\eeq
The magnitude of the numerator is
independent of the angle $\psi$ up to terms of order $R^4/h^4$.
This result shows that although particles are accelerated by the
Earth's magnetic field in the EW direction regardless of angle $\psi$,
the radiation received at the antenna does not show preference for the EW
polarization. Instead, it is directly related to the angle $\psi$.  As the
particle trajectory bends in the Earth's magnetic field and
the velocity deflects from the vertical direction, the
relation~(\ref{eqn:Eparallel}) between
the direction of the electric field vector and angle $\psi$ does not hold.
Nonetheless, it will be useful for understanding the angular dependence of
the electric field.

\begin{figure}[t]
\centerline{\includegraphics[width=0.75\textwidth]{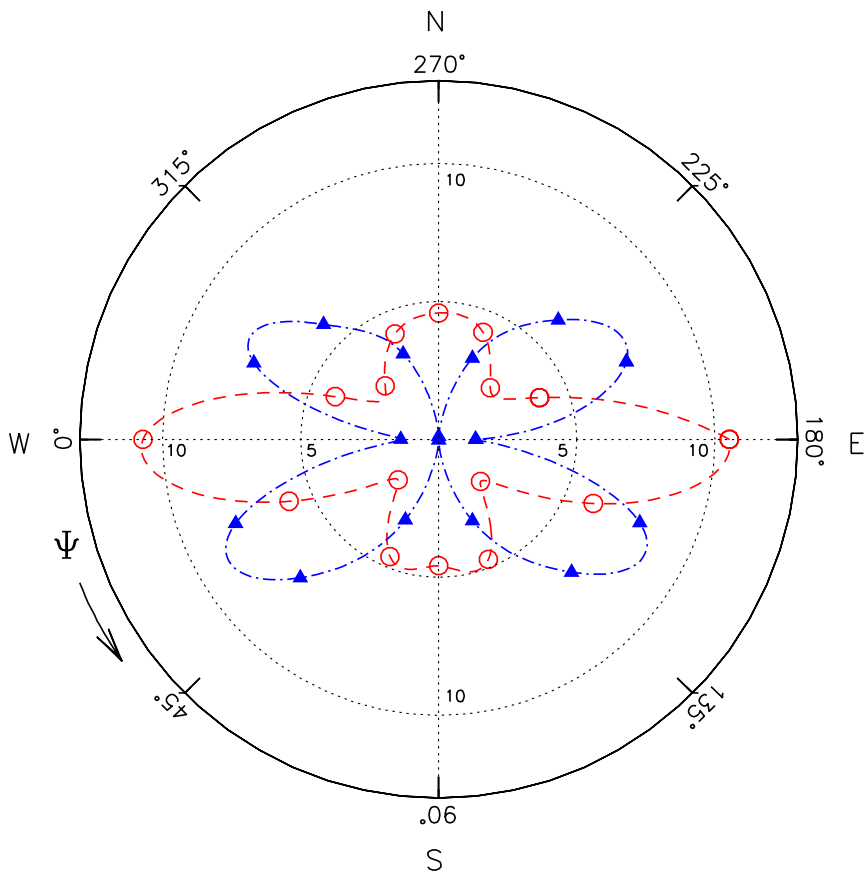}}
\caption{The East-West and North-South components of the field strength
$|{\cal E}_{\nu EW}|$ and $|{\cal E}_{\nu NS}|$ (circles and triangles,
respectively) at 55~MHz as functions of angle $\psi$ between the magnetic
West and direction to the shower core. The distance between the origin and a
circle or a triangle represents the field strength in the units of 
$\mu$V/m/MHz. The angular spacing between circles or triangles is 
$\pi/8$. At $\psi=\pm\pi/2$ $|{\cal E}_{\nu NS}|$ do not 
exceed~0.1~$\mu$V/m/MHz and two triangles overlap.
All points were calculated for the vertical shower at a 200~m distance 
from the antenna.}
\label{fig:EWNS}
\end{figure}

We computed electromagnetic pulses for the pancakes with axes located at
the same distance $R=200$~m from the antenna but at various angles $\psi$
from the $Ox$ direction. Fig.~\ref{fig:EWNS} shows the radio
signal strengths that would be received by EW and NS-oriented antennas.
Note that Eq.~(\ref{eqn:Eparallel}) predicts that components of the radiation
coming from the start of the particle trajectory vanish at some angles 
$\psi$: $E_{EW}=0$ at
$\psi=\pm\pi/4,\,\pm3\pi/4$, while $E_{NS}=0$ at $\psi=0,\,\pm\pi/2,\,\pi$.
This fact explains why ${\cal E}_{\nu EW}$ is relatively small at
$\psi=\pm\pi/4,\,\pm3\pi/4$ and ${\cal E}_{\nu NS}$ is
small at $\psi=0,\,\pi$ (Fig.~\ref{fig:EWNS}).
Another mechanism is responsible for ${\cal E}_{\nu NS}$ being virtually $0$ 
at $\psi=\pm\pi/2$. At these angles the trajectories of two charged particles
of an electron-positron pair are symmetric with respect to the $yOz$ plane. 
As we show in the Appendix the NS component of radiation emitted by this pair
vanishes not only at the start but throughout its flight. 

The unusually high value of ${\cal E}_{\nu EW}$ at $\psi=0$ and
$\pi$ can be explained when one takes a closer look at how the
denominator of the second term of Eq.~(\ref{elfield}) changes as
charged particles travel towards the Earth. As a positron starting
from $(R,\psi=0,h)$ in cylindrical coordinates bends to the East,
its velocity vector $\boldsymbol{\beta}$ becomes increasingly
closer to the vector ${\bf n}$ towards the antenna. As a result,
$ 1-n\boldsymbol{\beta}\cdot{\bf n} $ becomes smaller, the antenna
gets closer to the Cherenkov cone defined as
$ 1-n\boldsymbol{\beta}\cdot{\bf n} =0$ and the amplitude of the
radio signal increases.

The same mechanism accounts for the fact that the signal at
$\psi=\pi/8$ is larger than at $\psi=-\pi/8$. While bending around
the Earth's magnetic field ${\bf B}$, the originally vertical
velocities of the charged particles do not just acquire an
$x$-component. As a result of a nonzero dip angle $\gamma_d$ of
the ${\bf B}$ field, both electron and positron velocities also
get a small $y$-component towards the magnetic North as can be
seen from Eqs.~(\ref{eqn:ve}) and (\ref{eqn:vp}) below. This
brings the velocity vector $\boldsymbol{\beta}$ of the positron
starting at $\psi=\pi/8$ closer to the vector ${\bf n}$ towards
the antenna. For the positron at $\psi=-\pi/8$ the situation is
reversed. The North component of its velocity makes the angle
between the velocity and direction to the antenna larger than it
could have been without it. The third power in the denominator of
the Eq.~(\ref{elfield}) amplifies the effect and the difference
between signal amplitudes for these two angles becomes noticeable.

The symmetry of the plots with respect to the $Oy$ axis is remarkably good; 
the deviations between the corresponding values do not exceed 5\%. Also, the
field strengths ${\cal E}_{\nu NS}$ at $\psi=\pm\pi/2$ are very small
compared to those at other angles. These results tend to improve with a
higher statistics. The deviations between the symmetric dots become smaller
and the NS components of the field strengths at $\psi=\pm\pi/2$ decrease
even further. So, the deviations must be primarily statistical.
Both features are in full accord with theoretical predictions (see Appendix).

\section{Conclusions}

In this paper we presented a simple model of the cosmic ray air shower
radio emission based on calculation of the trajectories and radiation of the
charged particles of the shower maximum. Despite the limited nature of the
model, it correctly predicts the order of magnitude of the radiation
strength. We conclude that full scale shower development simulations
can be appended with calculations of radiation from individual shower
particles to give reliable predictions of the radiation properties.
If the Auger arrays are equipped with antennas to detect the shower
RF pulses, one would be able to check the model predictions for the
showers with energies higher than $10^{20}$~eV.

As one can see from Fig.~\ref{fig:EWNS}, the EW and NS components of the
radio pulse
are of the same order of magnitude and, hence, two perpendicularly oriented
antennas should be deployed at future detector sites to get a full
picture of shower radiation.
The prominent superiority of the EW component over the NS one at
$\psi=0,\,\pm\pi/2,\,\pi$ is quite sensitive to the value of the angle
$\psi$. A future detector should accumulate a large sample of
(almost) vertical showers before one would get radiation data from a few
showers with cores directly to the North, South, East or West of the antenna.
Then one should be able to confirm the conspicuous  dominance of the EW
component of their radiation over the NS one.

The polarization of the electromagnetic radiation is perpendicular to the
direction of its propagation.
For $10^{17}$~eV or weaker vertical showers with maximum mostly overhead, 
this means that the vertical component of the radiation is expected to be 
small. The relative strength of the vertical component is not small for 
inclined showers. Just like vertical showers, they radiate mostly in the
direction of their development and a perpendicular to this direction may
have an appreciable vertical component. However, inclined $10^{17}$~eV
showers travel through a greater atmospheric thickness before they reach 
the ground and develop higher 
above the Earth. The overall strength of their radiation should be small.
A greater penetration of very energetic ($\sim10^{20}$~eV) inclined showers
may compensate for this. In this case the shower radiation becomes stronger
and the vertical component may become detectable. An additional vertically
polarized antenna would facilitate the full measurement of the radiation from
these most intense air showers.

Fig.~\ref{fig:FT} shows a strong dependence of pulse
amplitude on the frequency. The optimal observation frequency band should be
chosen
with its lower limit at the lowest possible frequency consistent
with a relatively small RF
background (both natural and man-made), at around 20 or 30~MHz.

\section*{Appendix: Theory of electric field dependence on angle $\psi$}

In the frame of Fig.~\ref{fig:geometry} the Earth's magnetic field vector
${\bf B}$ is parallel to the vector $(0,-\cos\gamma_d,-\sin\gamma_d)$, where
$\gamma_d$ is the dip angle. The velocity vectors of electrons and positrons
bend around it in opposite directions. A straightforward calculation
gives the following expressions for position, velocity and acceleration of
an electron:
$$
{\bf r}^e(t)=
\left[R\cos\psi+\frac{\beta c \cos\gamma_d}{\omega}\, (1-\cos\omega t)
,\,
R\sin\psi-\frac{\beta c \sin\gamma_d\cos\gamma_d}{\omega}\,
\left(\omega t-\sin\omega t\right) \right.
,
$$
\beq
\left.
h-\beta ct+ \frac{\beta c \cos^2\gamma_d}{\omega} \,
\left(\omega t - \sin\omega t\right) \right] \ ,
\label{eqn:re}
\eeq
\beq
{\bf v}^e(t)=\beta c \ \left[\cos\gamma_d\sin\omega t,\,
-\sin\gamma_d\cos\gamma_d\,(1-\cos\omega t),\,
-1+\cos^2\gamma_d\,(1-\cos\omega t)\right] \ ,
\label{eqn:ve}
\eeq
\beq
{\bf a}^e(t)=\omega \beta c \cos\gamma_d \ (\cos\omega t,\,
-\sin\gamma_d\sin\omega t,\,\cos\gamma_d\sin\omega t) \ ,
\label{eqn:ae}
\eeq
where
\beq
\omega =\frac{eB}{\gamma m}
\eeq
is the gyration frequency and $t$ is the time in the observer's 
frame. $t=0$ corresponds to the start of the particle motion. 
If the index of refraction $n$ were constant through the atmosphere depth,
the relationship between the retarded time $t$ and the time $t_a$ when the 
signal arrives at the antenna would be $t_a=t+n\,|{\bf r}^e(t)|/c$. 
Time delays were corrected to take into account variations of the index of 
refraction with the altitude.

Similarly for a positron,
$$
{\bf r}^p(t)=
\left[R\cos\psi-\frac{\beta c \cos\gamma_d}{\omega}\,(1-\cos\omega t)
,\,
R\sin\psi-\frac{\beta c \sin\gamma_d\cos\gamma_d}{\omega}\,
\left(\omega t-\sin\omega t\right)  \right.
,
$$
\beq
\left.
h-\beta ct+ \frac{\beta c \cos^2\gamma_d}{\omega} \,
\left(\omega t -\sin\omega t\right) \right] \ ,
\label{eqn:rp}
\eeq
\beq
{\bf v}^p(t)=\beta c \ \left[-\cos\gamma_d\sin\omega t,\,
-\sin\gamma_d\cos\gamma_d\,(1-\cos\omega t),\,
-1+\cos^2\gamma_d\,(1-\cos\omega t)\right] \ ,
\label{eqn:vp}
\eeq
\beq
{\bf a}^p(t)=\omega \beta c \cos\gamma_d \ (-\cos\omega t,\,
-\sin\gamma_d\sin\omega t,\,\cos\gamma_d\sin\omega t).
\label{eqn:ap}
\eeq

The $|{\bf E}|$ independence of $\psi$ at $t=0$~\cite{NIM} does not hold as
electron-positron pairs travel towards the Earth. However, the symmetry
does not break down completely. We will show below that the symmetry with
respect to the $yOz$ plane remains, i.e., a pair starting from a point
$(R,\psi,h)$ in cylindrical coordinates produces an electric field of the
same magnitude as a symmetric pair from $(R,\pi-\psi,h)$.

First, note that a particle's velocity and acceleration are obviously
independent of $\psi$. For any time $t$ Eqs.~(\ref{eqn:ve}), (\ref{eqn:ae}),
 (\ref{eqn:vp}),  (\ref{eqn:ap}) give
\beq \beta^e_x=-\beta^p_x, \ \ \ \ \ \ \ \ \ \ \
\dot{\beta}^e_x=-\dot{\beta}^p_x, \label{eqn:first} \eeq \beq
\beta^e_y=\beta^p_y, \ \ \ \ \ \ \ \ \ \ \
\dot{\beta}^e_y=\dot{\beta}^p_y, \eeq \beq \beta^e_z=\beta^p_z, \
\ \ \ \ \ \ \ \ \ \ \dot{\beta}^e_z=\dot{\beta}^p_z. \eeq Now,
taking into account that ${\bf n}=-{\bf r}/r$, one can deduce from
Eqs.~(\ref{eqn:re}) and (\ref{eqn:rp}) that for any time $t$ \beq
(n_x)^e_\psi=-(n_x)^p_{\pi-\psi}, \eeq \beq
(n_y)^e_\psi=(n_y)^p_{\pi-\psi}, \eeq \beq
(n_z)^e_\psi=(n_z)^p_{\pi-\psi}. \label{eqn:last} \eeq The above
equations result in $(\boldsymbol{\beta}\cdot{\bf n})^e_\psi=
(\boldsymbol{\beta}\cdot{\bf n})^p_{\pi-\psi}$, and the
denominator of the second term in Eq.~(\ref{elfield}) is ``CP
symmetric". As for the numerator, we rewrite the double vector
product as ${\bf n}\,(\dot{\boldsymbol{\beta}}\cdot{\bf n})-
n\boldsymbol{\beta}\,(\dot{\boldsymbol{\beta}}\cdot{\bf
n})-\dot{\boldsymbol{\beta}}\, (1-n\boldsymbol{\beta}\cdot{\bf
n})$. The expressions in parentheses are ``CP symmetric". As for
vectors ${\bf n}$, $\boldsymbol{\beta}$ and
$\dot{\boldsymbol{\beta}}$, the
Eqs.~(\ref{eqn:first}-\ref{eqn:last}) tell that their $y$ and
$z$-components are ``CP symmetric", while the $x$-components are
antisymmetric. Finally, take into account the opposite charge
signs of positrons and electrons and obtain the following
relations between the components of the electric field vectors:
\beq (E_x)^e_\psi=(E_x)^p_{\pi-\psi}, \eeq \beq
(E_y)^e_\psi=-(E_y)^p_{\pi-\psi}, \eeq \beq
(E_z)^e_\psi=-(E_z)^p_{\pi-\psi}. \eeq

These relations can be directly translated to the case of two pancakes.
Pancakes with centers at $(R,\psi,h)$ and $(R,\pi-\psi,h)$ contain
symmetrically located electron-positron pairs. The $x$-components of the
 electric fields radiated by the pancakes and received at the antenna are
 the same and the $y$ and $z$-components only differ in sign. Thus, one
 expects a symmetry of the magnitude of any electric field component with
 respect to the $yOz$ plane.

One special case is a pancake at $\psi=\pi/2$ or $-\pi/2$. Each electron
inside this pancake has a counterpart -- a symmetrically located positron.
This implies that $y$ and $z$-components of the pancake's electric field
vanish.

\end{document}